\documentclass[10pt,conference]{IEEEtran}

\usepackage[pdftex]{graphicx}
\usepackage[utf8]{inputenc}

\graphicspath{{../figures/}}

\usepackage[nolist]{acronym}
\usepackage{url}

\usepackage{algorithm}
\usepackage{algorithmic}
\usepackage{calc}
\usepackage{amsmath}

\usepackage{amssymb}
\usepackage{amsfonts}
\usepackage{hyperref}
\usepackage{url}

\usepackage{xcolor}
\ifCLASSOPTIONcompsoc
    \usepackage[caption=false, font=normalsize, labelfont=sf, textfont=sf]{subfig}
\else
\usepackage[caption=false, font=footnotesize]{subfig}
\fi

\usepackage[multiuser,nomargin,marginclue,inline]{fixme}
\fxusetheme{color}
\FXRegisterAuthor{all}{anall}{ALL}
\FXRegisterAuthor{tolja}{antolja}{Tolja}
\FXRegisterAuthor{mc}{anmc}{Mikołaj}
\FXRegisterAuthor{awo}{anawo}{AWO}
\FXRegisterAuthor{pg}{anpg}{Piotr}

\newcommand{\fig}[1]{\figurename~\ref{#1}}

\fxsetup{status=draft}
\IEEEoverridecommandlockouts\IEEEpubid{\makebox[\columnwidth]{978-1-5090-0223-8/16/\$31.00 \copyright 2016 European Union \hfill} \hspace{\columnsep}\makebox[\columnwidth]{ }}
\begin{document}

\title{Towards efficient coexistence of IEEE~802.15.4e~TSCH and IEEE~802.11\\
\thanks{This work has been supported by the European Union’s Horizon 2020 research and innovation programme under grant agreement No. 645274 (WiSHFUL project).}
}

\author{
\IEEEauthorblockN{Mikołaj Chwalisz}
\IEEEauthorblockA{\textit{Telecommunication Networks Group (TKN)} \\
\textit{Technische Universität Berlin (TUB)}\\
Berlin, Germany \\
chwalisz@tkn.tu-berlin.de}
\and
\IEEEauthorblockN{Adam Wolisz}
\IEEEauthorblockA{\textit{Telecommunication Networks Group (TKN)} \\
\textit{Technische Universität Berlin (TUB)}\\
Berlin, Germany \\
wolisz@tkn.tu-berlin.de}
}

\maketitle

\begin{abstract}

A major challenge in wide deployment of smart wireless devices, using different technologies and sharing the same 2.4 GHz spectrum, is to achieve coexistence across multiple technologies.
The IEEE~802.11 (WLAN) and the IEEE 802.15.4e \acs{TSCH} (WSN) where designed with different goals in mind and both play important roles for respective applications.
However, they cause mutual interference and degraded performance while operating in the same space.
To improve this situation we propose an approach to enable a cooperative control which type of network is transmitting at given time, frequency and place.

We recognize that \ac{TSCH} based sensor network is expected to occupy only small share of time, and that the nodes are by design tightly synchronized.
We develop mechanism enabling over-the-air synchronization of the Wi-Fi network to the \ac{TSCH} based sensor network.
Finally, we show that Wi-Fi network can avoid transmitting in the "collision periods".
We provide full design and show prototype implementation based on the \ac{COTS} devices.
Our solution does not require changes in any of the standards.
\end{abstract}

\begin{IEEEkeywords}
Wireless LAN , Wireless sensor networks, Heterogeneous networks, Cooperative communication
\end{IEEEkeywords}

\section{Introduction}
\label{s:introduction}

The \ac{ISM} spectrum of 2.4 GHz is a very busy populated frequency spectrum that is \emph{intended} to be shared among numerous wireless technologies.
A very popular, if not the dominating technology is the omnipresent IEEE~802.11 (or Wi-Fi).
It is by design primarily optimized for high throughput, and features distributed self tuning without need of centralized management.
On the other hand, the increased interest in \acp{CPS} caused increased popularity of many \acp{WSN} technologies, such as WirelessHART, Zigbee, Bluetooth LE etc.
All of which also share prominent 2.4 GHz spectrum bandwidth.
The design of those technologies has been primarily driven by demands for long life thus energy efficiency, relatively low throughput rates and discontinuous data traffic.
Devices using these communication technologies are commonly battery powered and expected to operate over long periods (up to years) without maintenance.
The demand for low transmission power and limited communication ranges is, however, a disadvantage and vulnerability if they have to co-exist with other transmitters, which are often much stronger and can impair communications.

It is foreseeable that this different co-located wireless technologies will have to co-exist.
Unfortunately, this leads to cross-technology interference and degraded performance.
This work is motivated by a demanding version of the above scenario.
In particular we consider coexistence of two prominent technologies that target very different traffic types and are technically very different in nature.
Namely, the popular WLAN technology, IEEE~802.11 Wi-Fi~\cite{ieee80211-2012}, used for high throughput data streams.
And the IEEE~802.15.4e~\cite{ieee802154e-2012} using \acf{TSCH} mode of operation, which is gaining momentum in both standardization and deployment and constitutes a technology for highly reliable control networks.

Such combination will likely become typical for laboratories, and industrial settings, where a plant infrastructure is run over a highly reliable and latency constraint \ac{TSCH} network, while Wi-Fi networks run in parallel to serve high throughput multimedia data.
The challenge for coexistence of these technologies results from following features.
Operation of the TSCH nodes is strictly pre-scheduled, so they do not consider possible other activities as for per packet transmission. The only (limited) attempt for coexistence consists in avoiding frequency bands which seem heavily used by other systems. Unfortunate TSCH is reliable primarily due to frequency hopping, thus its operation is reasonable only if enough frequency bands are used.
Albeit Wi-Fi follows the listen before talk principle, it is unlikely to work properly for detection of TSCH transmissions.
The activity of low-power TSCH nodes is likely to be "overlooked" by Wi-Fi's carrier sensing based on signal power level.

\section{System Model}
\label{s:system_model}

\begin{figure}[h]
    \centering
    \includegraphics[width=0.6\columnwidth]{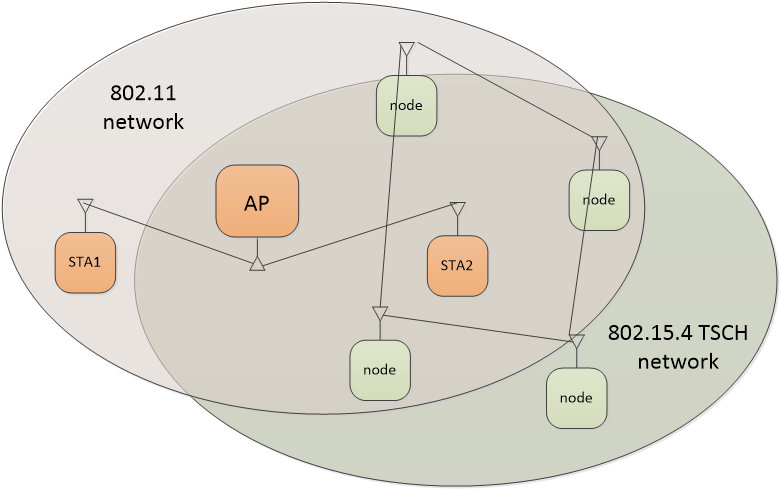}
    \caption{Two co-located wireless networks of different technology}
    \label{f:system_model}
\end{figure}

In this paper we consider initially a simplistic configuration of two networks:

\begin{itemize}
    \item
A single Wi-Fi basic service set working in the infrastructure mode i.e., consisting of one \acf{AP} and multiple Stations.
Wi-Fi devices under consideration are using \ac{COTS} hardware with the standard spectrum sensing capabilities.
Those devices are working on one channel in the $2.4$ GHz spectrum band.

    \item
A \ac{TSCH} mesh network operating in a frequency hopping scheduled mode in a way fully compliant to the standard~\cite{ieee802154e-2012} and using default parameter values if not specified otherwise.
No assumptions about specific features of the schedule to be used are made.
Thus, all 16 of the IEEE~802.15.4 channels might be used.
Note, that TSCH nodes do not support \acf{CCA} mechanisms.
\end{itemize}

Both networks are overlapping in space, as shown in \fig{f:system_model}.
They are working in the common $2.4GHz$ ISM frequency band, and thus interfere with each other under normal circumstances.
We assume that both of the above described networks have an Internet connectivity, and that it is possible to transmit some information among management process of the \ac{TSCH} network and the Wi-Fi management process within the \ac{AP}.

Petrova et al. in~\cite{petrova2007colocatedwifizigbee} showed, that "in an environment with a middle or high IEEE~802.11 traffic load it is very difficult to guarantee the quality of the nearby operating IEEE 802.15.4 based sensor networks".
Pollin et al. shows in experimental study~\cite{Pollin2008} the significant performance degradation of IEEE~802.11 network under the IEEE~802.15.4 interference.
In order to minimize interference, mechanisms to coordinate operation of both networks are needed.

In this paper we present, elaborate and test the following approach.
Recognizing that TSCH transmissions will occupy a small share of time, we claim that Wi-Fi network, if only correctly knowing when a collocated TSCH is going to transmit, can without significant capacity decrease, avoid transmitting in the "collision periods".
The key enabler is in the synchronization across both networks.
For that we present the full system design, as well as results of tests on operational prototype implemented using \ac{COTS} Wi-Fi nodes equipped with \emph{ath9k} based IEEE~802.11 wireless chipset, and TinyOS based fully TSCH compatible wireless sensor nodes~\cite{Watteyne2016}.
Having that it will be in general possible to control which type of network (or each device) is allowed to transmit in the given time, frequency and space.
As there is no possibility to directly (over the air) exchange data between cross-technology devices, it is necessary to acquire the \ac{TSCH} link schedule through a separate control channel.
We have evaluated the performance of the proposed solution.

\section{Related Work}
\label{s:related_work}

Huang et al. in~\cite{Huang2010} introduce a new scheme for allowing Zigbee links to achieve assured performance in the presence of heavy Wi-fi interference.
First, based on Wi-Fi traces they try to characterize the white spaces created in Wi-Fi traffic.
Based on that they create frame control protocol that limits the collision probability.
This work does not consider \ac{TSCH} standard and thus is not helpful in time scheduled sensor networks.

Li et al. in~\cite{Li2015a} present an adaptive TSCH channel selection scheme to improve reliability of the sensor network.
They argue that pseudo random hopping scheme of TSCH with static channel list is not enough to cope with interference generated by nearby Wi-Fi network.
In the presented solution each channel is selected independently, with a selection algorithm based on the multi-arm bandit problem.
The NS-3 simulation based results present $x1.5$ throughput improvement as compared to static and default channel list.

Gürsu et al. in~\cite{Gursu2016} present a way to improve TSCH packet error ratio by limiting the number of channels used for the hopping pattern.
They have showed decrease of packet drop rate while using only 4 out of 16 available channels in the evaluated setup.
Those channels where selected to minimize interference from Wi-Fi networks on channels 1,6,11.
Presented result is promising but does not fully solve the problem, as it violates the purpose of TSCH channel hopping.
By limiting number of channels it makes sensor network more prone to other interference.

Ruckebusch et al. in \cite{Ruckebusch2016} presented a cross technology synchronization using a prepared pattern send by a Wi-Fi node and received by sensor nodes.
They used custom Wi-Fi  and sensor node platforms to synchronize and create over arching time schedule between both networks.
This approach requires modification on all nodes, additionally senor nodes are not \ac{TSCH} compliant.

\section{\acs{TSCH} Primer}
\label{s:tsch_primer}

\acf{TSCH} is one of the MAC-modes defined in the IEEE-802.15.4e amendment \cite{ieee802154e-2012}.
It combines different well-known concepts of time, frequency and space diversity to meet today's request for reliable and latency constraint industrial networks.
The meshed \ac{TSCH} nodes are tightly synchronized to reduce energy consumption and perform channel hopping to increase reliability.
The standard defines mechanisms how scheduled network should operate, but by design it does not determine how schedule should be created.
The 6TisCH protocol stack defines the architecture and information needed to build the schedule.
Watteyne et al. in~\cite{Watteyne2016} describe current state of the standardization in that direction and present efforts in the four open-source implementations of these standards.

Time in \ac{TSCH} is slotted, and assumed to be (almost) perfectly synchronized in the whole system.
The basic time interval is referred to as a \acf{TS}.
It is sufficient for a transmission of one packet, including optional acknowledgment, between two neighboring nodes.
\ac{TSCH} uses a \ac{TS} counter called \acf{ASN} to determine the usage of the following \ac{TS}.
When a new network is created, the \ac{ASN} is initialized (usually to 0) and continues incrementing for each \ac{TS}, no matter if a given node is in this slot idle or involved in a transmission.
All nodes in a TSCH network have the same notion of \ac{ASN}.

A \emph{link} is the fundamental abstraction for describing the communication in \ac{TSCH}.
A link $l$ is defined as a tuple of transmitting node address $n$, receiving node address $d$, slot offset $s_o$, channel offset $c_o$ and link options $o$.
All the links defined for a given \ac{TSCH} system create a set $\mathbb{L}$ defining its operation.
\begin{equation}
l_i = \{n_{i}, d_{i}, s_{o,i},c_{o,i}, o_{i}\} \in \mathbb{L}
\end{equation}

Individual nodes only have knowledge of links that they participate in (either as transmitter or receiver).

Activity periods exceeding a single slot are structured in \acfp{SF}.
Each link has only one active slot within a \ac{SF} it belongs to.
Each \ac{SF} is characterized by its size $S$, defined as number of \acp{TS} composing this frame.
The standard does not impose a SF size.
Depending on the application needs (and platform capabilities), these can range from a few to 1000’s of \acp{TS}.
The shorter the SF, the more often a link has an active slot, resulting in more available bandwidth per link, but also in a higher power consumption for both transmitter and receiver on that link.
Multiple SFs can exist in parallel, which means that two links on one node but belonging to two different \ac{SF} can collide in time, such collisions are resolved by setting \ac{SF} priorities.

To increase the system reliability each link uses different channel for each transmission.
Possible channels are defined by the MAC hopping sequence list $\mathbb{C}$.
\begin{equation}
\mathbb{C} = [f_0, \dotsc, f_C], f_i \in \{11, \dotsc, 26\}
\end{equation}
The whole hopping network cannot communicate correctly unless it agrees upon the hopping sequence $\mathbb{C}$ being used.
It can be either pre-configured or changed dynamically and announced in beacons.

The frequency channel $c$ used by a given node at given time is calculated dynamically.
For a slot with \ac{ASN} $a$ and given channel offset $c_o$ it is calculated with:
\begin{equation}
c(a,c_o) = \mathbb{C}[(a + c_o)\bmod C]
\label{e:tsch_channel}
\end{equation}

Note, that the given link $l_i$ is active only once per \ac{SF}.
It will happen only when:
\begin{equation}
s_{o,i} = a \bmod S
\end{equation}

The logical representation of the activity of all links in the network is called a \emph{schedule}.
Schedules are mostly computed in a centralized way, albeit distributed schemata are also considered.
In this work we are not considering the schedule creation, we assume that a schedule is computed and remains valid for a long time period.
Additionally, we consider schedules with single \acf{SF}.
This constraint simplifies the investigation but doesn't limit the applicability of the solution.
For more detailed information about \ac{TSCH} refer to~\cite{ieee802154e-2012, Watteyne2015}.

\section{Cross technology synchronization}
\label{s:synchronization}

The key problem that needs to be solved is getting two wireless systems synchronized, so that both systems can perform operations using the same time reference.
We assume knowledge of the \ac{TSCH} schedule consisting of description of all links, and slot frame parameters as described in Sec.~\ref{s:tsch_primer}.
This information can be easily transfered over the backhaul Internet connection, but it doesn't contain any timing reference.
Achieving tight time synchronization over the backhaul channel introduces a lot of requirements and problems in deployment.
There is also no way of direct over-the-air communication, due to the physical layer incompatibility.

\subsection{Approach}

First, consider the energy pattern generated by a \ac{TSCH} network in the frequency spectrum.
For a particular link $l_i$ we define a signal $\Gamma_i[a]$ as a function of \ac{ASN}.
The value of $\Gamma_i[a]$ is the channel number used for the transmission if the link $l_i$ is active and $0$ otherwise.
\begin{equation}
\Gamma_i[a] = \begin{cases}
    \mathbb{C}[(a + c_i) \bmod C],& \text{if } s_{o,i} = a \bmod S\\
    0,              & \text{otherwise}
\end{cases}
\end{equation}

From that we can express a 2D binary signal $\Psi_i[a,c]$ for a link $l_i$ at a slot $a$ and channel $c$, marked with $1$ if there is a transmission expected on this channel and time and $0$ otherwise.
\begin{equation}
\Psi_i[a,c] = \begin{cases}
    1,& \text{if } \Gamma_i[a] = c\\
    0,              & \text{otherwise}
\end{cases}
\end{equation}

Furthermore, we can compute signal $\Psi[a,c]$ for all links, which is defined as $1$ if any of the links are active and $0$ otherwise.

\begin{equation}
\Psi[a, c] = 1- \prod_{l_i \in L} (1 - \Psi_i[a,c])
\label{e:tsch_signal}
\end{equation}

Considering schedule information transfered over backhaul channel, we can treat $\Psi[a,c]$ as a known signal.
The channel hopping mechanism makes it resemble a random signal.
This is a desired property of a \ac{TSCH} network, as it provides statistically good interference avoidance.

We will use the following well known approach to detect the known signal $\Psi[a,c]$ by performing spectrum sensing $P[a,c]$, and calculating cross-correlation.
\begin{align}
(\Psi \ast P)[m] &= \sum_{a=-\infty}^{\infty}\sum_{c \in \mathbb{C}} \Psi[a,c] P[a+m,c] \\
                 &= \sum_{a=-\infty}^{\infty}\sum_{c \in \mathbb{C}} \Psi[a,c](\Psi[a+m,c] + N[a+m,c])
\end{align}
Lag with maximum value will show time shift between generated and measured signal, thus giving synchronization information.
For other lags, the cross-correlation is expected to be small because $\Psi[a,c]$ is close to random.
It is important to note that we perform cross-correlation only in time (slot) dimension as the channels used are already known.
In practice it would be necessary to perform cross-correlation for an arbitrarily big lags to find proper match.
This limits applicability of the idea in real systems.

To overcome this problem we can observe existence of the intrinsic unique and finite pattern in $\Psi[a,c]$.
Even though the schedule repeats with total number of time slots in the \ac{SF}, the resulting computed sequence of channels for the \acp{TS} repeats with a lower frequency.
This is caused by the way the pseudo-random channel hopping is computed.
The modulus operator in the channel $C$ calculation~\eqref{e:tsch_channel} causes the repeating \ac{SF} to use different channels as in previous iterations.
The full period $\lambda$ depends on the lengths of the \ac{SF} and channel list.
It is given by~\eqref{e:schedule_periodicity}, which is a least common multiple of all \ac{SF} lengths $S$ and a hopping sequence list length $C$.
It is expressed in number of slots, but can be converted to Seconds by multiplying with slot duration ($10 ms$ by default).

\begin{equation}
\lambda = S \sqcup C \\
         = \frac{S \cdot C}{S \sqcap C}
\label{e:schedule_periodicity}
\end{equation}

As a result the auto-correlation of a periodic signal is also periodic with the same period, preserving the property of having low values at lags other than the period.
This reduces the required correlation time to the period lengths.

\begin{equation}
(\Psi \ast P)[m] = \sum_{a=0}^{\lambda}\sum_{c \in \mathbb{C}} \Psi[a,c] P[a+m,c]
\end{equation}

\begin{figure}[htb]
\centering
\includegraphics[width=\linewidth]{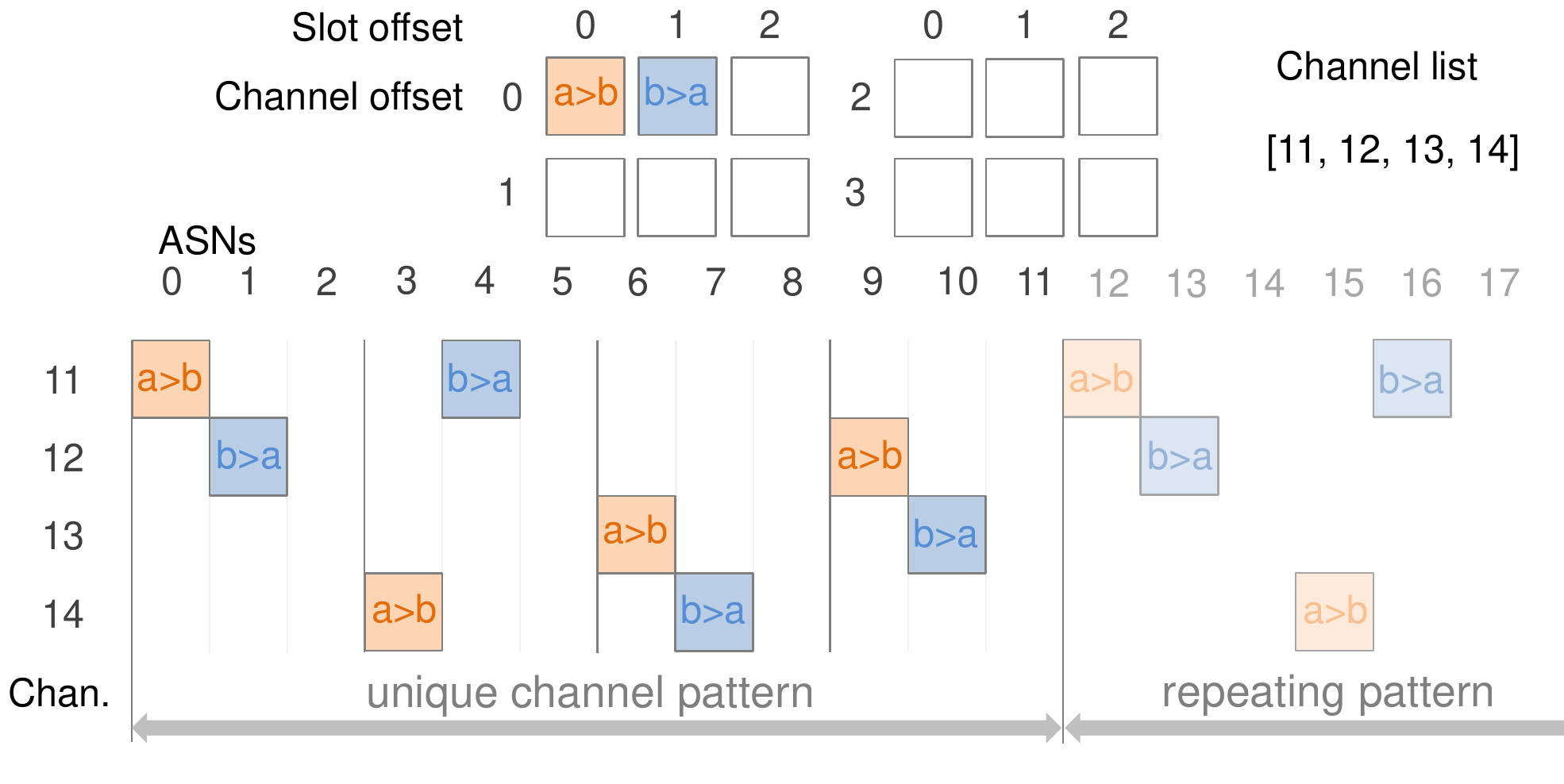}
\caption{TSCH Schedule}
\label{f:simpleschedule}
\end{figure}

As an example, \fig{f:simpleschedule} shows a schedule containing only one \ac{SF},
It is three slots long, has defined two links, two nodes are participating in the network, both of the links are using channel offset $0$.
The channel list consists of four channels.
We can observe that the whole signal will repeat after twelve slots.
Note, the links although having channel offset $0$ are using different channels on consecutive \ac{SF} repetitions.
It is possible to generate signal $\Psi[a,c]$ for any given schedule.

\subsection{Spectrum sensing}
\label{s:ath_sensing}

To achieve over-the-air cross-technology time synchronization we first need to detect \ac{TSCH} signal $\Psi[a,c]$ and use it to extract necessary timing information.
In other words we are detecting a distinct event happening in the TSCH network and time stamping it with the clock from sensing device.
We can use the spectral samples collected from the wireless chipsets used as a sensing device.

Atheros chipsets, based on \emph{ath9k} or \emph{ath10k} Linux drivers, include a built-in spectral analysis features~\cite{ath9kspectral}.
They are able to report FFT data from the baseband, including the absolute magnitude for each of 56 FFT bins\footnote{128 FFT bins in case of \texttt{HT40} mode} in a given Wi-Fi channel.
The user has an ability to control how often the spectral data is reported to the kernel and through \emph{debugfs} interface to the user process.
The samples are time stamped with the accurate \ac{TSF} value~\cite{Mahmood2014, Mahmood2014a}, that can be used later to extract timing information for a whole Wi-Fi \ac{BSS}.

It is important to note that there are several limitations to the \emph{ath9k} spectral scanning functionality.
First, and most important, the chipset will not report any spectral information while in transmit or receive mode.
That includes also periods when card is in receive mode but decoding yields \ac{FCS} errors.
Second, the samples are gathered opportunistically, which in other words means that the timing between samples is not fixed.
This needs to be addressed, as the most of the cross-correlation techniques require uniform distribution of samples and the same sampling rate in both data sets.

We setup \emph{ath9k} device to collect spectrum measurements continuously (in \texttt{background} mode) and process data in batches.
We decouple data collection and further processing.
The sensing process should be parametrized so as to assure the synchronization quality without introducing too much data processing overhead.
We setup one batch of collected data to be at least two $\lambda$ period durations (in Seconds), to be sure it is possible to detect one full period of signal $\Psi[a,c]$.
We have configured the system to receive spectral measurements every $1ms$, which is an order of magnitude faster than the default \ac{TSCH} slot size of $10ms$.

In the pre-processing step we need to re-sample raw measurements to be sampled every exactly $1ms$, taking the maximum for each time bin at each frequency bin.
We introduce zero samples for missing data, that appear due to Wi-Fi transmissions.
As the result, we obtain signal $P[t,f]$ represented as a function of time $t$ in \emph{Seconds} and frequency bin $f$ in \emph{Hertz}.

\subsection{\ac{TSCH} spectral model}
\label{s:tsch_model}

To cross-correlate measurements $P[t,f]$ with model $\Psi[a,c]$ it is necessary to represent both signals in the same units.
Next, we will show the procedure to represent $\Psi[a,c]$ as function of time and frequency.
Discrete slots need to be converted into time values, while the channel numbers into frequency values.
Computed model has to cover $\lambda$ slots to cover one full period of the signal $\Psi$.
This has to be done once for a given schedule and has to match resolution of the $P[t,f]$ signal.
Both conversions are orthogonal and can be interchanged.

\paragraph*{Time domain}
The $\Psi[a,c]$ signal needs to be resampled into the same rate as the spectral scan ($1ms$).
The new $\Psi[t,c]$ signal is set to $0$ when slot is not active at given time and channel.
We observe that not all portions of active \acp{TS} are being used for transmissions.
Therefore, we limit the fraction of time the signal is active to resemble the transmission characteristics of a slot.
Based on~\cite{ieee802.15.4-2015}, we set to $1$ only actual maximal transmit duration of \emph{macTsMaxTx} ($4256\mu\textrm{s}$) of each active slot after silent offset \emph{macTsTxOffset} ($2120\mu\textrm{s}$).
The remainder of the slot is also set to $0$, which ignores acknowledgments in the first approximation.

\begin{figure}[h]
\centering
\includegraphics[width=.7\columnwidth]{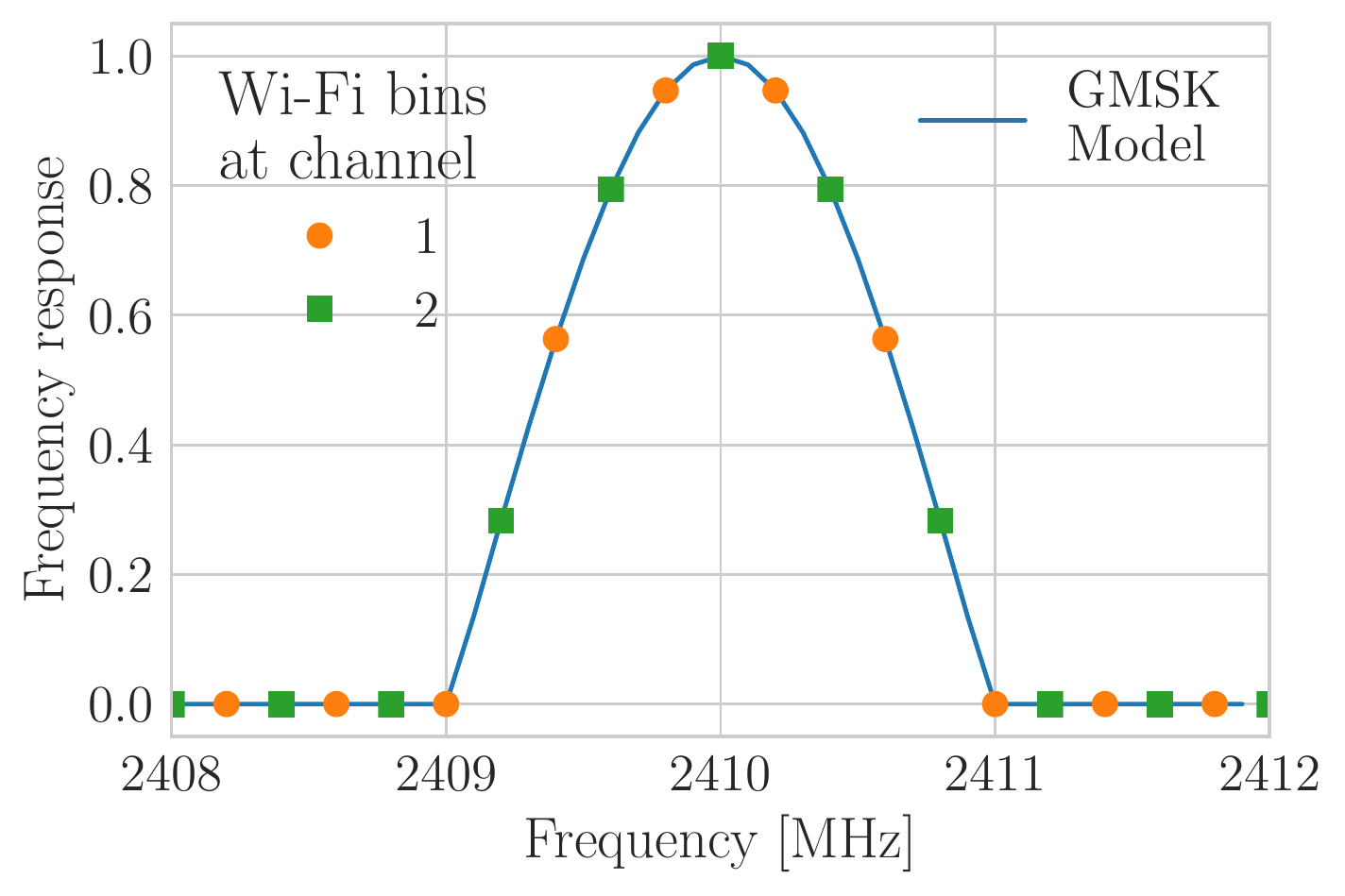}
\caption{IEEE~802.15.4 spectrum model}
\label{f:channel_model}
\end{figure}

\paragraph*{Frequency domain}
Knowing that IEEE~802.15.4 PHY layer uses \ac{GMSK} modulation~\cite{ieee802.15.4-2015}, we can model power spectral density.
\fig{f:channel_model} shows the generated model for channel 12 (.15.4).
The power is scaled between $0$ and $1$ as only relative values matter for cross-correlation.
Such solution allows for any given IEEE~802.15.4 channel to be generated for any frequency list, and thus it is possible to work with arbitrary spectrum sensing data.
In particular, the spectrum view from the \emph{ath9k} chipset is interesting.
Depending on the relative difference between channels and fixed FFT binning different parts of sensor network channel can be observed by Wi-Fi device.
This can be seen in~\fig{f:channel_model} in the mapping of \emph{ath9k} FFT bins seen from Wi-Fi channels 1 and 2.

For each active sample in $\Psi[t,c]$ signal we generate a model for channel $c$ and frequency list used by \emph{ath9k} spectrum sensing.
IEEE~802.15.4 channels not overlapping with the current Wi-Fi channel can be discarded in the further consideration, as they cannot be detected.
Inactive samples are converted to zero array for the given frequency list.
Sum of all samples yields a desired $\Psi[t,f]$ signal.

\paragraph*{Example}

\begin{figure}[ht]
  \centering
  \subfloat[Example schedule]{\includegraphics[width=1\columnwidth]{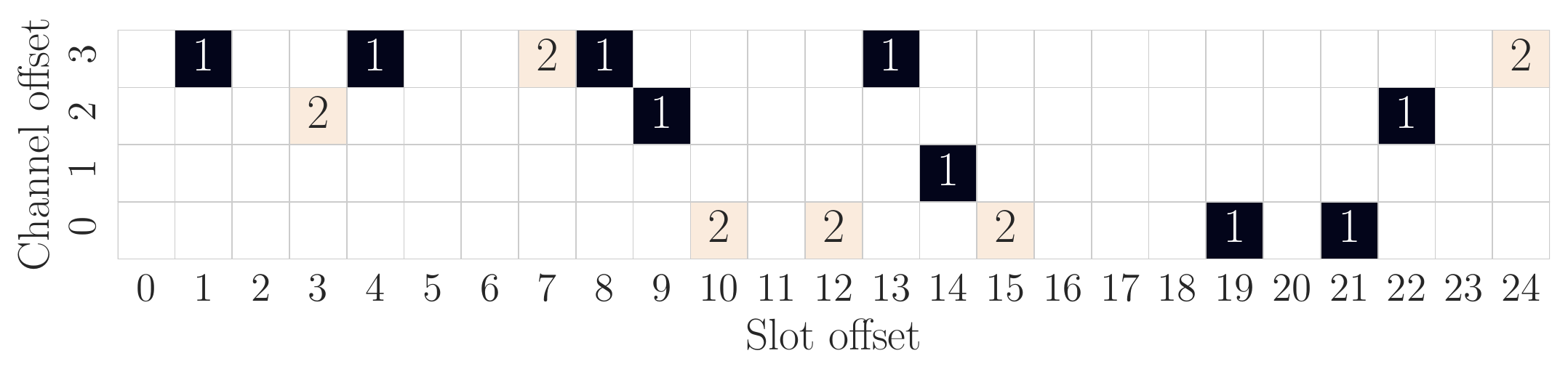}
    \label{f:olaf}\hfill}
  \qquad
  \subfloat[Generated model]{\includegraphics[width=0.9\columnwidth]{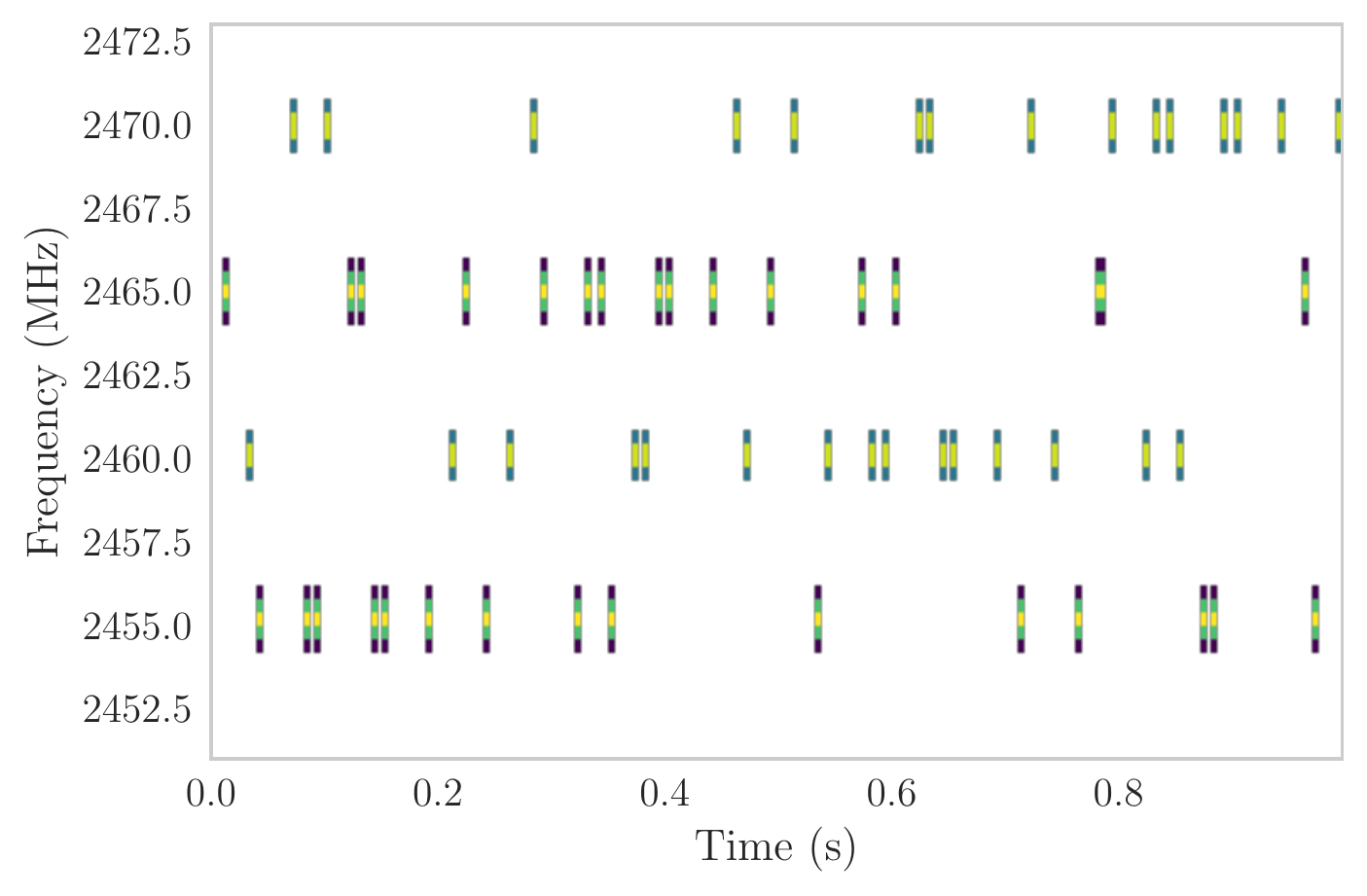}
    \label{f:olaf_model}\hfill}
  \caption{\acs{TSCH} schedule model}
  \label{tf:olaf}
\end{figure}

Taking the schedule presented in \fig{f:olaf}, where the transmitting node ID is shown for active slots.
Knowing the \ac{TSCH} network is using channels $[11, 12, 13, 14]$ and Wi-Fi network channel $11$.
It is possible to generate model shown in \fig{f:olaf_model}.
This schedule will be also used in evaluation described in Section~\ref{s:evaluation}.

This procedure can be adapted to any sensing device.
Only the conversion parameters need to be adapted based on the sensing device capabilities.

\subsection{Cross correlation}
\label{s:cross_correlation}

We define a time reference signal as a point of time when the \ac{TSCH} network starts new repetition of a pattern $\Psi[a,c]$.
The generated model $\Psi[t,f]$ and measured $P[t,f]$ can be used to find when this event occurs.
We correlate both signals $(\Psi \ast P)[m]$ over time dimension using normalized cross-correlation method presented in~\cite{Lewis1995}.
The lags $m$ for the highest peaks are representing the points of time for the searched reference.
Those are the times where there is a best match between model and measurements.
This allows for getting a mapping between the two clocks and thus achieving synchronization.

We treat the discovered signal as statistically significant if the value of the cross-correlation at lag $m$ is bigger than 3 times standard deviation from the mean of the whole cross-correlation result.
Given lags can to be converted back to the \ac{TSF} timestamps acquired from the measured signal.
Those timestamps can be used in Wi-Fi network for cooperation with \ac{TSCH} network.
It is possible to detect it in retrospect with couple of seconds delay and calculate mapping for the TSF based clock.

This synchronization signal is valid for the whole \acf{BSS}, as \ac{TSF} counters are kept synchronized (via beacons) in all stations.
This also means, it is enough to perform sensing and synchronization on one station, and distribute detected \ac{TSF} value to other nodes to achieve Wi-Fi and \ac{TSCH} synchronization throughout both networks.
Sensing can be performed on any station in the \ac{BSS}, not necessarily at the \ac{AP}.
It is enough to disseminate the \ac{TSF} counter value to all other stations.
It is possible to acquire synchronization signal in retrospect, with couple of seconds delay, and calculate next events in the \ac{TSF} based clock.

\section{Wi-Fi cooperation}
\label{s:cross_tdma}

The synchronization between \ac{TSCH} and Wi-Fi network can be achieved by performing spectrum sensing by any node in \ac{BSS}.
Good candidates are \acp{AP}, as they play a central role in the Wi-Fi network, or nodes that are known to be close to \ac{TSCH} network.
The synchronization information can be used to control the behavior of the Wi-Fi nodes.
Assuming, low utilization of the \ac{TSCH} network it makes sense to avoid Wi-Fi transmissions in times when there are a scheduled \ac{TSCH} node transmissions overlapping in frequency.
There are three problems that need to be addressed.
First, how to control Wi-Fi nodes to prevent transmissions in the designated times.
Additional constraint is to avoid changes to the IEEE~802.11 standards.
Second, assuming the \ac{TSCH} schedule is known, it is necessary to calculate when Wi-Fi nodes should cease transmission.
Third, limiting overhead of acquiring synchronization, how to disseminate synchronization signal to all nodes in the \ac{BSS}.

To configure behavior of the Wi-Fi nodes we base the approach on the hybrid medium access architecture presented by Zehl et al. in~\cite{zehl16hhmac}.
The solution exploits the 802.11 power saving functionality to enable control of the software packet queues in the Linux driver for the Atheros chipsets.
As the result it is possible to control when the wireless interface will transmit packets over the air, this pertains to \ac{AP} as well as all stations.
Note, this solution doesn't enable power saving mode of the card, it allows for disabling software packet queues and thus preventing any packet being pushed to the chipset.

We are using the proposed and modified \emph{ath9k} kernel driver to control the devices and un/pause software queues.
We have extended the PyRIC\footnote{\url{https://github.com/wraith-wireless/PyRIC}} Python library to support this new kernel functionality.
With such addition, and with usage of persistent \emph{Netlink} sockets, it is possible to achieve fast interface from control process running in Python to the \emph{ath9k} kernel module.
Still, whole control is performed in user- not in kernel-space.

The proposed solution suffers from the same limitation as~\cite{zehl16hhmac} in terms of the accuracy in the \emph{Netlink} communication.
Namely, every change of state needs to be actively initiated by the user-space application.
Additionally, it is necessary to convert TSF timestamps to the UNIX host clock.
This can lead to the additional inaccuracies and delays in the execution.

To correctly control Wi-Fi nodes using proposed approach, it is necessary to provide slotting information i.e., points in time when transmission queues are should be un/paused.
First, from $\Psi[a,c]$ we take four IEEE~802.15.4 channels\footnote{There will be 8 channels in case of \texttt{HT40} mode.} that are overlapping with currently used Wi-Fi channel.
Wi-Fi channel should be reserved, for \ac{TSCH} network, when at least one of the sensor node channels are active for a given slot.
This gives the binary array of slots being reserved or free for Wi-Fi use.
Interesting are only transitions of state, as only then there needs to be control action performed.
When channel becomes free at the beginning of the slot the Wi-Fi queues should be immediately unpaused, allowing for Wi-Fi communication.
On the other hand, to ensure free channel for scheduled \ac{TSCH} transmission, Wi-Fi chipset needs to be able to transmit all remaining packets before channel will be used by sensor nodes.
Knowing that the \ac{TSCH} node will not transmit until \emph{macTsTxOffset} ($2120 \mu\textrm{s}$).
Taking, 6Mbit/s as the slowest bit-rate for Wi-Fi and the max packet size of 1500B, it takes $2158 \mu\textrm{s}$ to transmit it with \ac{ACK}.
From the above we can to conclude that the software queues need to be blocked at least $38 \mu\textrm{s}$ before \ac{TSCH} slot start, this does not yet include synchronization inaccuracy.

To achieve the best isolation between both networks, all Wi-Fi nodes in \ac{BSS} should cease transmissions in case of \ac{TSCH} communication.
This means all Wi-Fi nodes in the \ac{BSS} should perform local control of the packet queues.
Fortunately, all can use the same slotting and synchronization information, as the local \ac{TSF} counters are kept synchronized by the Wi-Fi devices.
The slotting information is expected to only change sporadically and the synchronization information when new sensing is available, assuming newer information is better because it reduces clock drift.
Both can be transmitted with normal Wi-Fi network.

\section{Evaluation}
\label{s:evaluation}

We build on Watteyne et al.~\cite{Watteyne2016} work concerning the implementation of the \ac{TSCH} standard.
We have been using NXP JN5168 sensor nodes using TinyOS based standard compliant \ac{TSCH} implementation.
Wi-Fi devices are TP-Link WDR4300 \acp{AP} and Intel NUC PCs, all equipped with Atheros based wireless cards.
The spectrum sensing is performed on Intel NUC PCs.

The prototype implementation of the system was done in Python using ZeroMQ~\cite{Hintjens2012} for communication between processing elements.
Such design allows for decoupling between spectrum sensing, data processing, synchronization extraction and Wi-Fi cooperation control.
This enables to run each on different host without additional setup.
The local control of the Wi-Fi packet queues runs on each Wi-Fi node, all subscribe necessary synchronization information via ZeroMQ socket.
It is possible to perform on-line and off-line data processing by storing raw sensing data to disk.

\begin{figure}[ht]
\centering
\includegraphics[width=\columnwidth]{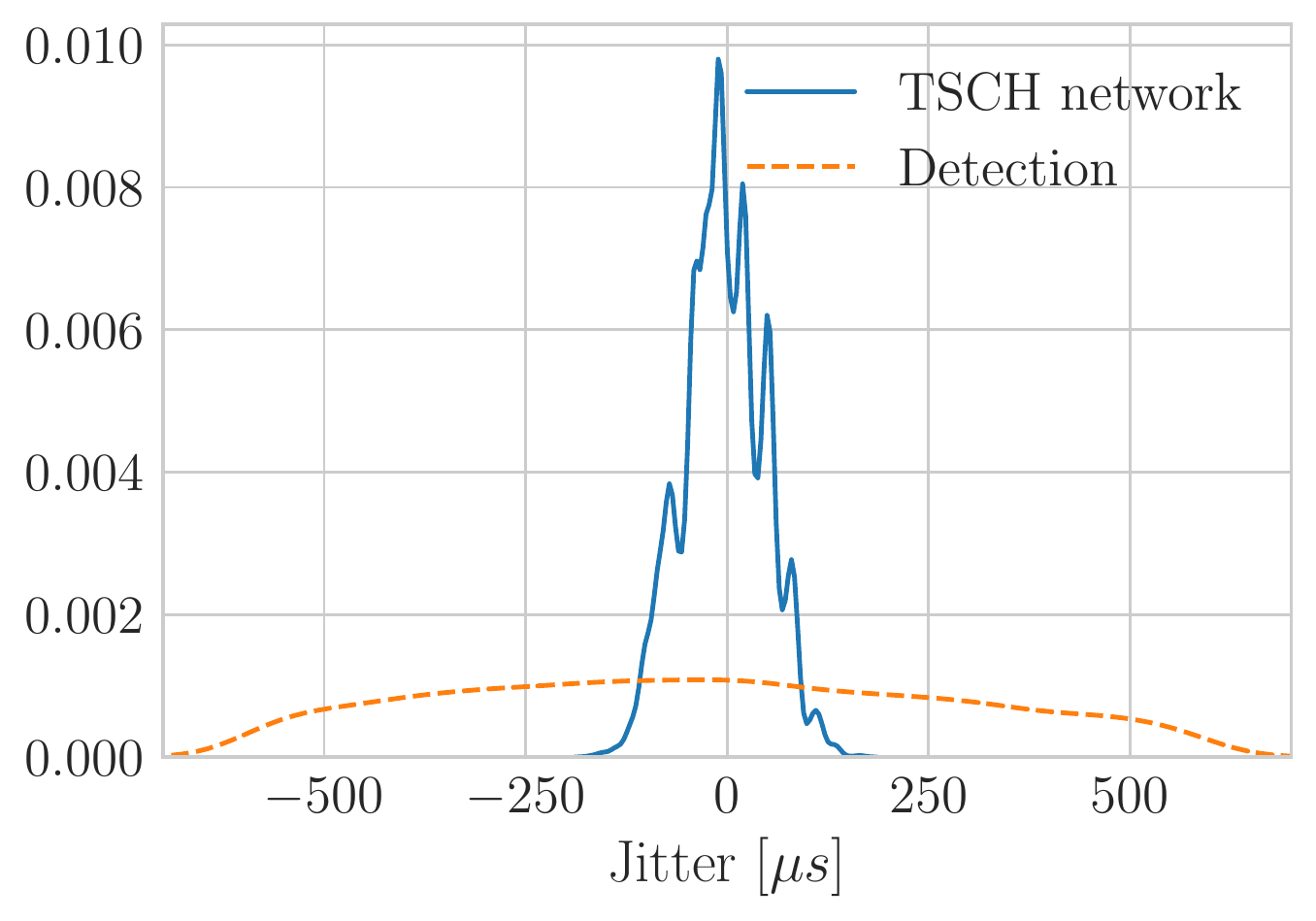}
\caption{Jitter of TSCH clock}
\label{f:jitter_detected}
\end{figure}

The \ac{TSCH} network is using schedule presented in~\fig{tf:olaf}.
Synchronization signal coming from sensor node was tracked using GPIO pins connected to high accuracy GPS based timing device.
At the same spectrum was sensed by Intel NUC PC and the time reference was computed by the presented algorithm.
\fig{f:jitter_detected} shows the kernel density estimate of synchronization jitter as computed from both sources.
Note, for the detected time reference the difference of consecutive detections where calculated, some events are not detected.
Each correlation was checked if the detected peak is significant, i.e. larger than 3 standard deviations from correlation mean, data was ignored otherwise.

The standard deviation of the \ac{TSCH} network clock jitter is $49\mu s$, while for the detected time reference it is $310 \mu s$.
This values need to be taken into account as an additional guard times for Wi-Fi slotting.

The proposed sensing solution causes no overhead on the \ac{TSCH} network, which is important for sensor node battery life.
There is sensing and data processing overhead on one of the Wi-Fi nodes, which is in current prototype implementation significant.
Constant on-line data processing uses around $50\%$ of the Intel i5-4250U CPU.
This overhead can be greatly reduced by doing spectrum sensing less often, e.g. every couple of seconds and by improving implementation.

\section{Conclusions}

The interference between devices is the key problem in the current wireless networks.
Improving the situation is particularly difficult if we consider simultaneous operation of incompatible technologies like IEEE~802.11 and IEEE~802.15.4e.
By assuming only limited out-of-band communication between such networks, we have shown a cross technology synchronization algorithm that allows a Wi-Fi network to acquire time reference signals from running TSCH network.
This time reference signal can be used by a Wi-Fi network to keep the channel free for the time of TSCH transmissions and use regular (CSMA based) transmission in during rest of the time.
All this can be achieved on \ac{COTS} devices, without modifying technology standards they operate in, which is a major benefit for bringing this technology into real usage.

We currently assume a policy in which WLAN traffic is less critical and can tolerate delays.
There is a need to consider the utilization of the \ac{TSCH} network, as very busy sensor network can render Wi-Fi unusable.
Future network control policies should consider a trade off between utilization of both networks and set priorities accordingly.

Additionally, there is still an open question of suitable algorithms for assigning slots in the \ac{TSCH} networks.
Strategies favoring cooperation between networks should be investigated and promoted.
For example, assigning four slots, with channel offsets set to perform four simultaneous transmissions in consecutive channels, could occupy only one time slot for Wi-Fi network.
It is also interesting to investigate auto-correlation properties of the \ac{TSCH} schedule, in order to improve discoverability.

We have assumed that both networks are of similar size and thus fully in an interference range.
Considering one Wi-Fi \ac{BSS} and a bigger a \ac{TSCH} network, e.g. covering a whole building or industrial plant.
If is easy to imagine a situation where only part of such sensor network will be impacted by one \ac{AP} (with clients).
In such case, the sensing and synchronization quality might drop, because only subset of \ac{TSCH} nodes will be sensed by a Wi-Fi node.
This problem can be mitigated by performing multiple cross-correlations with different \ac{TSCH} network models, taking different subset of nodes for each.
Result with highest peak to mean ratio of cross-correlation function will be the best candidate for synchronization signal.

The more challenging problem is making a decision which nodes need to be protected from Wi-Fi communication i.e., when Wi-Fi nodes should cease communication.
Safe but inefficient is protection of the whole \ac{TSCH} network communication, even knowing that some nodes are not impacted by given Wi-Fi \ac{BSS}.
Such approach can make Wi-Fi unusable, with large number of sensor nodes and artificially busy schedule.
Optimally, only the sensor nodes, which packet reception is impaired by Wi-Fi communication, should be protected.
Due to incompatible \aclp{PHY}, between Wi-Fi and sensor networks, we cannot assume reciprocity of communication or sensing between technologies.
Thus, without any additional information it is not possible to find optimal solution to this problem.
Additional protocols need to be investigated to help in such situations.

\bibliographystyle{IEEEtran}
\bibliography{001_XX+mendeley}

\begin{acronym}

\acro{COTS}{Commercial off-the-shelf}
\acro{CSMA}{Carrier Sense Multiple Access}
\acro{FFT}{Fast Fourier Transform}
\acro{GMSK}{Gaussian Minimum Shift Keying}
\acro{GPS}{Global Positioning System}
\acro{ISM}{Industrial, Scientific and Medical}
\acro{LAN}{Local Area Network}
\acro{MAC}{Medium Access Control}
\acro{PHY}{Physical Layer}
\acro{TDMA}{Time Division Multiple Access}
\acro{WLAN}{Wireless LAN}

\acro{ACK}{Acknowledgment}
\acro{CCA}{Clear Channel Assessment}
\acro{IE}{Information Element}

\acro{CPS}{Cyber-Physical System}
\acro{WSN}{Wireless Sensor Network}

\acro{ASN}{Absolute Slot Number}
\acro{EAck}{Enhanced Acknowledgment}
\acro{EB}{Enhanced Beacon}
\acro{SF}{Slot Frame}
\acro{TSCH}{Time-Slotted Channel Hopping}
\acro{TS}{Timeslot}

\acro{AP}{Access Point}
\acro{BSS}{Basic Service Set}
\acro{FCS}{Frame Check Sequence}
\acro{MPDU}{MAC protocol data unit}
\acro{STA}{Station}
\acro{TSF}{Timing Synchronization Function}

\acro{NTP}{Network Time Protocol}
\acro{PTP}{Precision Time Protocol}
\acro{PPS}{Pulse Per Second}

\end{acronym}

\end{document}